\documentclass[11pt]{article}

\voffset= - 1.1in
\hoffset= - 1.0in         

\makeatletter
\newdimen\normalarrayskip              
\newdimen\minarrayskip                 
\normalarrayskip\baselineskip
\minarrayskip\jot
\newif\ifold             \oldtrue            \def\new{\oldfalse}
\def\arraymode{\ifold\relax\else\displaystyle\fi} 
\def\eqnumphantom{\phantom{(\theequation)}}     
\def\@arrayskip{\ifold\baselineskip\z@\lineskip\z@
     \else
     \baselineskip\minarrayskip\lineskip2\minarrayskip\fi}
\def\@arrayclassz{\ifcase \@lastchclass \@acolampacol \or
\@ampacol \or \or \or \@addamp \or
   \@acolampacol \or \@firstampfalse \@acol \fi
\edef\@preamble{\@preamble
  \ifcase \@chnum
     \hfil$\relax\arraymode\@sharp$\hfil
     \or $\relax\arraymode\@sharp$\hfil
     \or \hfil$\relax\arraymode\@sharp$\fi}}
\def\@array[#1]#2{\setbox\@arstrutbox=\hbox{\vrule
     height\arraystretch \ht\strutbox
     depth\arraystretch \dp\strutbox
     width\z@}\@mkpream{#2}\edef\@preamble{\halign
\noexpand\@halignto
\bgroup \tabskip\z@ \@arstrut \@preamble \tabskip\z@ \cr}%
\let\@startpbox\@@startpbox \let\@endpbox\@@endpbox
  \if #1t\vtop \else \if#1b\vbox \else \vcenter \fi\fi
  \bgroup \let\par\relax
  \let\@sharp##\let\protect\relax
  \@arrayskip\@preamble}
\def\eqnarray{\stepcounter{equation}%
              \let\@currentlabel=\theequation
              \global\@eqnswtrue
              \global\@eqcnt\z@
              \tabskip\@centering
              \let\\=\@eqncr
 \halign to \displaywidth\bgroup
    \eqnumphantom\@eqnsel\hskip\@centering
    $\displaystyle \tabskip\z@ {##}$%
    \global\@eqcnt\@ne \hskip 2\arraycolsep
         $\displaystyle\arraymode{##}$\hfil
    \global\@eqcnt\tw@ \hskip 2\arraycolsep
         $\displaystyle\tabskip\z@{##}$\hfil
         \tabskip\@centering
    &{##}\tabskip\z@\cr}
\begingroup\ifx\undefined\newsymbol \else\def\input#1 {\endgroup}\fi

\def\beq{\begin{equation}}
\def\eeq{\end{equation}}
\def\ba{\beq\new\begin{array}{c}}
\def\ea{\end{array}\eeq}
\def\be{\ba}
\def\ee{\ea}

\def\N2{${\cal N}=2$}

\def\c{\check}
\def\p{\partial}

\newcommand{\Tr}{\mathop{\rm Tr}\nolimits}

\newcommand\vol{\mathop{\rm Vol} \nolimits}

\title{{\bf
Solving Virasoro Constraints
in Matrix Models}
\vspace{.5cm}}
\author{{\bf A.Alexandrov}\thanks{E-mail:
\ al@itep.ru}
\date{ } \\
{\small {\it MIPT} and
{\it ITEP, Moscow, Russia}}\\ \\
{\bf A.Mironov}\thanks{E-mail:
\ mironov@itep.ru; mironov@lpi.ac.ru}
\date{ } \\
{\small {\it Theory Department, Lebedev Physics Institute}
and {\it ITEP, Moscow, Russia}}\\ \\
{\bf A.Morozov}\thanks{E-mail: \ morozov@itep.ru}
\date{ } \\ {\small
{\it ITEP, Moscow}
}}

\begin{document}

\setcounter{footnote}{3}

\maketitle

\vspace{-9cm}

\begin{center}
\hfill FIAN/TD-15/04\\
\hfill ITEP/TH-49/04\\
\end{center}

\vspace{7.5cm}

\begin{abstract}
This is a brief review of recent progress
in constructing solutions to the matrix model
Virasoro equations. These equations are parameterized
by a degree $n$ polynomial $W_n(x)$, and the general solution
is labeled by an arbitrary function of $n-1$ coefficients of the
polynomial. We also discuss in this general
framework a special class of (multi-cut)
solutions recently studied in the context
of ${\cal N}=1$ supersymmetric gauge theories.
\end{abstract}

\vspace{1cm}

\paragraph{\large Introduction.}
It was realized in the beginning of nineties that matrix models partition
functions typically satisfy an infinite set of Virasoro-like equations
\cite{cVir,dVir}. These were nothing but Ward identities (Schwinger-Dyson
equations) which mainly fixed matrix model partition functions (because of
the topological nature of matrix models \cite{DijWit},
the Ward identities were restrictive
enough). Moreover, it turned out that one of the most technically
effective ways to deal with matrix models was to solve these Virasoro equations
(they are also sometimes called loop equations)
\cite{loop,K,Akemann,amm1,amm2}.

At early times of matrix models one usually dealt with Virasoro equations
describing relatively simple ``phases" so that the equations had unambiguous
solutions. An interest to more complicated phases of matrix models has
revived
after G.Bonnet, F.David and B.Eynard \cite{David} proposed to deal with the
multi-cut (or multi-support) solutions (known for long,
\cite{multicut,dav,Akemann, Ch}) in a new way: releasing the tunneling
constraint \cite{dav}. Their approach was later applied by
R.Dijkgraaf and C.Vafa \cite{DV} to
description of low energy superpotentials in ${\cal N}=1$
SUSY gauge theories,  \cite{CIV}.

More concretely, the authors of \cite{CIV} considered the ${\cal N}=2$ SUSY gauge
(Seiberg-Witten) theory in special points where some BPS states become
massless. Therefore, these states can condense in the vacuum which breaks
half of the supersymmetries (leading to ${\cal N}=1$ SUSY) and gives rise to
a non-zero superpotential. Values of this superpotential in minima are related
to the prepotential of a Seiberg-Witten-like theory. In turn, R.Dijkgraaf and
C.Vafa associated \cite{DV} the prepotential
with logarithm of a partition function of the Hermitean
one-matrix model in the planar limit of multi-cut solutions (it was later
proved in \cite{CM}).

In fact, actual definition of the multi-cut partition functions is a separate
problem.
For instance, one could simply define them as (arbitrary) solutions to the
corresponding Virasoro equations (D-module point of view). Then, one may ask
what is special about the concrete Dijkgraaf-Vafa (DV) solutions.
In the present paper we make further steps in this direction,
discuss the space of all solutions to the Virasoro
equations in the multi-cut phase and show that the DV
solutions form a basis in this space. They are distinguished by a special
property of isomonodromy that allows one to associate with these
Seiberg-Witten-like systems a Whitham hierarchy \cite{CM,CMMV}, the
corresponding partition function having a
multi-matrix model integral representation \cite{David,KMT}.

\paragraph{\large Hermitean one-matrix model.}

Hermitean one matrix model is given by the formal matrix integral over
$N\times N$ Hermitean matrix $M$
\be\label{maint}
Z_W(t)\equiv {1\over\hbox{Vol}_{U(N)}}\int DM \exp{1\over g}\left(
-\Tr W(M)+\Tr\sum_k t_kM^k\right)
\ee
Here $W(x)$ is an arbitrary function that we usually assume to be a
polynomial of degree $n+1$, $W_n(x)\equiv\sum_k^{n+1} T_kx^k$ and
the constant $g^2$ controls the genus expansion.
This integral still needs to be defined. One possibility is to substitute it
with its saddle point approximations
\cite{David,KMT}. Different saddle-points $M = M_0$ are given by the
equation $W'(M_0) = 0$. If the polynomial
\be
W'(x) = \prod_{i=1}^n(x-\alpha_i)
\ee
has roots $\alpha_i$, then, since $M_0$ are matrices
defined modulo $U(N)$-conjugations (which allow one
to diagonalize any matrix and permute its eigenvalues),
the different saddle points are represented by
\be
M_0 =  diag(\alpha_1,\ldots,\alpha_1;
\alpha_2,\ldots,\alpha_2;\ \ldots\ ; \alpha_n,\ldots,\alpha_n)
\ee
with $\alpha_i$ appearing $N_i$ times, $\sum_{i=1}^n N_i = N$.
In fact, there is no need to keep
these $N_i$ non-negative integers: in final expressions (like formulas
for the multidensities and prepotentials)
they can be substituted by any complex numbers. Moreover,
$N_i$ can depend on $T_k$ (i.e. on the shape of $W(\phi)$) and $g$.

Now, using at intermediate stage the eigenvalue representation of matrix
integrals, one can rewrite \cite{David,KMT} the matrix integral (\ref{maint})
over $N\times N$ matrix $M$ as $n$-matrix integral over $N_i\times N_i$
matrices $M_i$ (each obtained with the shift by $\alpha_i$: just changing
variables in the matrix integral (\ref{maint})), which is nothing but the DV
solution \cite{DV}
\be
Z_W^{(matr)}(t|M_0)\sim
\int \prod_{i=1}^n DM_i \exp\left(\sum_{i,k}\Tr t_k^{(i)}M^k \right)
\prod_{i<j}^n\alpha_{ij}^{2N_iN_j}\times\\\times
\exp \left(2\sum_{k,l=0}^\infty (-)^{k}\frac{(k+l-1)!}
{\alpha_{ij}^{k+l}k!l!}\Tr_iM_i^k\Tr_jM_j^l\right)
\ee
The variables $t^{(i)}_k$ are given by the relation
\be
\sum_{k=0}^\infty t_k \left(\sum_{i=1}^n \Tr_i(\alpha_i + M_i)^k\right) =
\sum_{i=1}^n \left(\sum_{k=0}^\infty t_k^{(i)} \Tr_i M_i^k\right)
\label{ttalpha}
\ee
with arbitrary $N_i\times N_i$ matrices $M_i$.

\paragraph{\large Virasoro constraints.}
The other possibility is to observe that (\ref{maint}) satisfies
the infinite set of (Virasoro) equations
(=Schwinger-Dyson equations,=Ward identities) \cite{dVir}
\be
\hat L_m Z_W(t) = 0, \ \ m\geq -1
\label{vir}\\
\hat L_m = \sum_{k\geq 0} k\left(t_k-T_k\right)
\frac{\partial}{\partial t_{k+m}} + g^2
\sum_{\stackrel{a+b=m}{a,b\geq 0}} \frac{\partial^2}{\partial t_a\partial t_b}
\ee
\be
{\p Z_W\over\p T_k}+{\p Z_W\over\p t_k}=0 \ \ \ \ \forall k=0,...,n+1
\ee
and call any solution to these equations the matrix model partition
function. Then, the partition function is not a function but
a formal $D$-module, i.e. the entire collection of power series
(in $t$-variables), satisfying a system of consistent linear
equations. Solution to the equations does not need to be unique, however,
an appropriate analytical continuation in $t$-variables transforms
one solution to another, and, on a large enough moduli space
(of {\it coupling constants} $t$), the whole entity can be considered,
at least, formally as a single object: this is what we call
{\it the partition function}.
Naively different solutions are interpreted as different {\it branches}
of the partition function, associated with different {\it phases} of the theory.
Further, solutions to the linear differential
equations can be often represented as integrals (over {\it spectral
varieties}), but integration ``contours'' remain unspecified: they can
be generic {\it chains} with complex coefficients (in the case of integer
coefficients this is often described in terms of {\it monodromies}, but
in the case of partition functions the coefficients are not restricted
to be integer).
{\it A model} of partition function is an integral formula which has
enough many free parameters to represent the generic solution of the
differential equations in question.

A familiar example that could clarify these notions is provided with
the cylindric functions. Their defining equation is
\be
\left[\lambda^2\partial_{\lambda}^2 +\lambda\partial_{\lambda}+
\left(\lambda^2-\nu^2 \right)\right]Z_{\nu}(\lambda)=0
\ee
and an integral representation is
\be\label{cylint}
Z_{\nu}(x)={1\over 2\pi}\int_C e^{-ix\sin\theta+i\nu\theta}d\theta
\ee
{\it The model} is given by the generic
linear combination of two contours, say,
chosen as in 8.423 of \cite{GR} (this choice fixes as the basis the Hankel
functions).

\paragraph{\large Loop equations.}
Another form of the Virasoro equations (the loop equation) is produced by
rewriting the infinite set of these equations through the
unique generating function of all single trace correlators
\be
\rho^{(1)}(z|t)\equiv
\hat\nabla (z) {\cal F},\ \ \ \ \hat\nabla (z)\equiv\sum_{k\ge 0}
{1\over z^{k+1}}{\p\over\p t_k},\ \ {\cal F}\equiv g^2\log Z_W
\ee
Introducing notations $v(z)$ for $\sum_k t_kz^k$ and $\left[...\right]_+$
( $\left[...\right]_-$) for the projector onto non-negative (negative)
degrees of $z$, one obtains {\it the loop equation} \cite{loop}
\be\label{loop}
W'(z)\rho^{(1)}(z|t)=
\left(\rho^{(1)}(z|t)\right)^2 + f_W(z|t) + g^2\hat\nabla(z)\rho^{(1)}(z|t) +
\left[v'(z)\rho^{(1)}(z|t)\right]_-
\ee
\be\label{f}
f_W(z|t)\equiv \left[W'(z)\rho^{(1)}(z|t)\right]_+\equiv\hat R (z) {\cal F}
\ee
In order to consider (connected) multi-trace correlators, one needs
to introduce higher generating functions (also named loop mean, resolvent etc)
which we call multi-density
\be\label{corrho}
\rho^{(m)}(z_1,...,z_m|t)\equiv\hat\nabla (z_1)...\hat\nabla (z_m){\cal F}
\ee
In what follows, we consider solutions to the Virasoro equations (\ref{vir})
as a formal series in $t$-variables, as well as a series in the coefficient
$g^2$ (genus expansion):
\be
\log Z_W=g^{-2}{\cal F}=\sum_{p\ge 0}g^{2p-2}{\cal F}^{(p)}
\ee

\bigskip

\framebox{\parbox{5.8in}{

\bigskip

\begin{center}
{\Large\bf \underline{Main results.}}
\end{center}

\bigskip

Here we are going to review briefly the main results of the papers \cite{amm1}
and \cite{amm2}, where we defined the matrix model partition function as any
solution to the Virasoro equations (\ref{vir}). In forthcoming paragraphs
we briefly comment on these results.

\begin{itemize}

\item Any solution to the Virasoro constraints (taken as a formal series in
$t$-variables and in genus expansion) is unambiguously labeled by an
arbitrary function of $n$ of $n+2$ $T$-variables: the {\it bare}
all genera prepotential.

\item There is {\bf an evolution operator} that generates from
the $t$-independent bare prepotential the matrix model partition function
which depends on
$t$-variables and satisfies the Virasoro equations. This evolution
operator {\it does not depend} on the choice of the arbitrary function, but
only on $T$- and $t$-variables.

\item One may invariantly define ``the occupation numbers" of
\cite{David,DV} as eigenvalues of operators constructed from the evolution
operator, formula (\ref{ocnum}) below. The corresponding DV solutions are
described as eigenfunctions of these operators.

\item ({\bf Conjecture 1}) The evolution operator can be completely expressed
in terms of the unique operator
\be\label{checkR}
\check y\equiv \sqrt{W'(x)^2 -4g^2 \c R(x)},\ \ \ \ \check R (x)\equiv
-\sum_{a,b=0}(a+b+2)T_{a+b+2}x^a{\p\over\p T_b}
\ee
its derivatives and $W'(x)$.

\item ({\bf Conjecture 2}) The evolution operator is constructed as a formal
series in $t$ with operator coefficients acting on the bare prepotential.
These coefficients generate the full matrix model
correlators. These operator coefficients are related to {\bf operator
multidensities} (\ref{corrho})
exactly as the full correlators are related to the
connected correlators, only an appropriate ordering prescription should be
applied. This relation {\bf is
universal}, i.e. is the same for the Gaussian (quadratic) and non-Gaussian
potentials.

\item ({\bf Conjecture 3}) The ordering used in the previous conjecture
is not uniquely defined.

\end{itemize}
}}

\paragraph{\large Solving the Virasoro constraints.}

In order to convert
(\ref{loop}) into a solvable set of recurrent relations, we expand
$\rho^{(1)}(z|t)$ in powers of $g^2$ and $t$'s
\be
\rho^{(1)}(z|t) = \sum_{p,m\geq 0} \frac{g^{2p}}{m!}
\oint \ldots \oint v(z_1)\ldots v(z_m) \rho^{(p|m+1)}(z,z_1,\ldots,z_m),
\\
f_W(z|t) = \sum_{p,m\geq 0} \frac{g^{2p}}{m!}
\oint \ldots \oint v(z_1)\ldots v(z_m)
f_W^{(p|m+1)}(z|z_1,\ldots,z_m)
\ee
In this way, we introduce the full set
of multidensities $\rho_W^{(p|m)}(z_1,\ldots,z_m)$
and auxiliary polynomials $f_W^{(p|m+1)}(z|z_1,\ldots,z_m)$
(which distinguishes between different phases for a given $W(z)$)
at zero $t$'s.

Acting on eq.(\ref{loop}) with the operator
$\hat\nabla(z_1)\ldots\hat\nabla(z_m)$ and putting all $t_k=0$ afterwards,
we obtain a double-recurrent (in $p$ and $m$) relation for
the multidensities $\rho^{(p|m)}$
\be
W'(z)\rho^{(p|m+1)}(z,z_1,\ldots,z_m) -
f_W^{(p|m+1)}(z|z_1,\ldots,z_m) = \\ =
\sum_q \sum_{m_1+m_2=m}
\rho^{(q|m_1+1)}(z,z_{i_1},\ldots,z_{i_{m_1}})
\rho^{(p-q|m_2+1)}(z,z_{j_1},\ldots,z_{j_{m_2}}) + \\ +
\sum_{i=1}^m \frac{\partial}{\partial z_i}
\frac{\rho^{(p|m)}(z,z_1,\ldots,\check z_i,\ldots,z_m) -
\rho^{(p|m)}(z_1,\ldots,z_m)}{z-z_i}
+
\hat\nabla(z)\rho^{(p-1|m+1)}(z,z_1,\ldots,z_m)
\label{recrel1}
\ee
Together with (\ref{f}) this relation is enough
to find explicit expressions for all the multidensities through
$W(z)$ and $f^{(p|1)}_W(z)$. In fact, the latter polynomials
(all of degree $n-1$) are not
independent, since for $m=0$,
\be\label{fthrF}
f^{(p|1)}_W(z) = \check R F^{(p)}[T]
\ee
expresses all the $f$-polynomials
through a single function of $T$ (i.e. of $W(z)$) and $g$,
the prepotential at $t=0$,
\be
{\cal F}(t=0,g) = F[g,T] = \sum_{p=0}^\infty g^{2p} F^{(p)}[T]\equiv
Z[g,T]
\ee
The operator $\check R$ here is given in (\ref{checkR}) and
contains derivatives with respect to the $T$-variables.
We call such operators {\it check operators} and denote
by the ``check" sign to distinguish them from {\it the hat operators}, which
contain $t$-derivatives and are denoted by ``hats".

Note that the $T_k$ dependence of $F[g,T]$ is not arbitrary,
since the first two ($\hat L_{-1}$ and $\hat L_0$) Virasoro constraints
are linear in derivatives and can be
consistently truncated to $t=0$ and then allow one to express two
derivatives, say, $\partial F/\partial T_{n+1}$ and
$\partial F/\partial T_{n}$ through $\partial F/\partial T_{l}$
with $l = 0,\ldots, n-1$. As a corollary,
the partition function can be represented as
\be
\left.Z(T)\right|_{t=0} = Z[g,T]=
\int dk z(k;\eta_2,\ldots,\eta_n;\hbar)
e^{\frac{1}{\hbar}(kx-k^2w)}
\label{etaparam}
\ee
with an arbitrary function $z$ of $n$ arguments $(k,\eta_2,\ldots,\eta_n)$
and $\hbar$. Here the $\hat L_{-1}$-invariant variables are used,
\be
w = \frac{1}{n+1}\log T_{n+1}, \ \ \
x  = T_0 + \ldots \sim \eta_{n+1},\\
\eta_k = \left(T_n^k + \frac{k(k-2)!}{n!}\sum_{l=1}^{k-1} (-)^l
\frac{(n+1)^l (n-l)!}{(k-l-1)!}
T_{n-l}T_n^{k-l-1}T_{n+1}^l
\right)T_{n+1}^{-\frac{kn}{n+1}}
\ee

As an immediate corollary of (\ref{recrel1}),
we obtain for $p=0$ and $m=0$
\be
\rho^{(0|1)}(z) = \frac{W'(z) - y(z)}{2}
\label{sdenW}
\ee
with
\be
y^2(z) = (W'(z))^2 - 4 f^{(0|1)}_W(z)
\ee

\paragraph{\large Evolution check operator.}

The basic claim is that, for any $W(z)$, there is an evolution
(check) operator $\check U_W(t)$, which converts {\it any} function
$Z[T]$ of $T_0,\ldots,T_{n-1}$
(with prescribed dependence on $T_n$ and $T_{n+1}$) into
$Z_W(t) = \check U_W(t)Z[T]$, which satisfies the Virasoro constraints,
$L_m(t)Z_W(t) = 0, \ \ m\geq -1$.
This means that the evolution operator is the same for any values
of the arbitrary parameters $f$ (or for any function $Z[T]$) once $W(z)$ is
fixed and that ``orbits" of the evolution operators are completely
parameterized by $W(z)$. Moreover, if
$Z[T] = \sum_a c_a Z^{(a)}[T]$, then $Z_W(t) = \sum_a c_aZ^{(a)}_W(t)$. This
means that one may arbitrarily choose a basis in the space of all solutions,
with the evolution not disturbing the expansion of any solution w.r.t.
this basis.

One may construct the operator $\check U_W(t)$ with the following procedure.
For given $T$'s, we make use of the Virasoro constraints
$\hat L_m Z(t) = 0$
and their multiple $t$-derivatives to recurrently express
$\left.\frac{\partial^p Z}{\partial t_{k_1}\ldots \partial
t_{k_p}}\right|_{t=0}$ with all $0 \leq k_i < \infty$ through
$g^{2s}\left.\frac{\partial^{p+s} Z}{\partial T_{l_1}\ldots \partial
T_{l_{p+s}}}\right|_{t=0}$ with all $0 \leq l_j < n$
\be
\left.\frac{\partial^p Z}{\partial t_{k_1}\ldots \partial
t_{k_p}}\right|_{t=0}
 = \sum_{\stackrel{s}{l_1,\ldots,l_{p+s}}}
g^{2s}{\cal D}^{l_1\ldots l_{p+s}}_{k_1\ldots k_p}
\left.\frac{\partial^{p+s} Z}{\partial T_{l_1}\ldots \partial
T_{l_{p+s}}}\right|_{t=0}
\ee
This is a straightforward procedure, and the sum over $s$ is finite,
from $0$ to the integer part of $\left\{\frac{k}{n-1}\right\}$:
the expression for
$\left.\frac{\partial Z}{\partial t_k}\right|_{t=0}$ contains
$\left.\frac{\partial^2 Z}{\partial t_a\partial t_b}\right|_{t=0}$,
but with $a,b \leq k-n-1$, further, the expression for
$\left.\frac{\partial^2 Z}{\partial t_a\partial t_b}\right|_{t=0}$
contains
$\left.\frac{\partial^3 Z}{\partial t_a\partial t_b\partial
t_c}\right|_{t=0}$,
this time with $a,b,c \leq k-2n-2$ and so on.

Let us now define the operators
\be
\check D^{(p)}_{k_1\ldots k_p} =
\sum_{\stackrel{s}{0\leq l_1,\ldots,l_{p+s} \leq n-1}}
g^{2s}{\cal D}^{l_1\ldots l_{p+s}}_{k_1\ldots k_p}
\left.\frac{\partial^{p+s}}{\partial T_{l_1}\ldots \partial
T_{l_{p+s}}}\right|_{t=0}
\ee
and construct the evolution operator $\check U_W(t)$ as a series in these
$\check D$-operators
\be
\check U_W(t) = 1 + t_k\check D_k^{(1)} +
\frac{1}{2}t_kt_l \check D_{kl}^{(2)} +
\frac{1}{6}t_kt_lt_m\check D_{klm}^{(3)} +
\ldots
\ee
The fact that, for any $Z[T]$,
\be
\hat L_m(t) Z_W(t) = \hat L_m(t)  \check U_W(t) Z[T] = 0
\ee
or, simply, that
\be
\hat L_m(t) \check U_W(t) =
\left(\sum_k kT_k \check D_{k+m}^{(1)} +
\sum_{a+b = m} \check D_{ab}^{(2)}\right)  +
\sum_l t_l \left(l\check D_{l+m}^{(1)} + \sum_k kT_k \check D_{k+m,l}^{(2)} +
\sum_{a+b = m} \check D_{abl}^{(3)}\right) +\\+
\frac{1}{2}\sum_{l_1,l_2} t_{l_1}t_{l_2}
\left(l_1\check D_{l_1+m,l_2}^{(2)} + l_2\check D_{l_2+m,l_1}^{(2)} +
\sum_k kT_k \check D_{k+m,l_1,l_2}^{(3)} +
\sum_{a+b = m} \check D_{abl_1l_2}^{(4)}\right) + \ldots =0
\label{lincom}
\ee
is equivalent to vanishing of all the linear combinations of operators
in brackets, and these are the characteristic equations for the
$\check D$-operators.

\paragraph{\large Basis in the space of all solutions.}

One can now cleverly choose some basis in the space of all solutions.
Note that the DV solutions do form such a basis. To have them written in a
more clever way, one may present the contribution of a particular extremum
$M_0$ (labeled by the set of $N_i$) above in the Givental-style
decomposition form,
expressing it through the product of $n$ Gaussian partition functions
($Z_G^{{\cal M}}(t|N)$ given by the $N\times N$ matrix integral with quadratic
$W_G(x)\equiv {\cal M}x^2$), with
its own $N_i$ and ${\cal M}_i = W^{\prime\prime}(\alpha_i) =
\prod_{j\neq i}(\alpha_i - \alpha_j)$,
\be
Z_{W}^{(matr)}(t|M_0) =
\frac{\prod_{i=1}^n e^{-N_iW(\alpha_i)}\vol_{U(N_i)}}{\vol_{U(N)}}
\prod_{i<j}^n\alpha_{ij}^{2N_iN_j}
\prod_{i<j}^n \hat{\cal O}_{ij} \prod_{i=1}^n
\hat{\cal O}_i \prod_{i=1}^n Z_G^{{\cal M}_i}(t^{(i)}|N_i)
\label{factfor}
\ee
with operators
\be
\hat{\cal O}_{ij} = \exp \left(2\sum_{k,l=0}^\infty (-)^{k}\frac{(k+l-1)!}
{\alpha_{ij}^{k+l}k!l!}
\frac{\partial}{\partial t^{(i)}_k}\frac{\partial}{\partial t^{(j)}_l}
\right),
\ee
$\alpha_{ij} = \alpha_i-\alpha_j$, and
\be
\hat{\cal O}_i = \exp \left(-\sum_{k\geq 3} \frac{W^{(k)}(\alpha_i)}{k!}
\frac{\partial}{\partial t^{(i)}_k}\right)
\ee
($W^{(k)}(x) = \partial_x^kW(x)$).

These DV solutions have a series of properties that basically have
much to do with isomonodromic deformations (by $W(x)$) and are in charge of
Whitham integrable systems behind the DV solutions \cite{CM,CMMV}, see the
end of the next paragraph.

\paragraph{\large Evolution operator as a function of $\check y$.}

So far we basically considered the ``connected" correlation functions,
$\rho^{(\cdot|m)} (z_1,\ldots,z_m;g)$. The other possibility is to deal with
``full" correlation functions,
\be
K^{(\cdot|m)}(z_1,\ldots,z_m;g) =
\left.
Z_W(t;g)^{-1}\hat\nabla(z_1)\ldots \hat\nabla(z_m)Z_W(t;g)
\right|_{t=0} =
\sum_{p=0}^\infty
g^{2(p-m)}
K^{(p|m)}(z_1,\ldots,z_m)
\label{corK}
\ee
These are related by
\be
K^{(\cdot|m)}(z_1,\ldots,z_m;g) =\\=
\sum_\sigma^{m!} \ \sum_{k=1}^{m} \
\sum_{\nu_1,\ldots,\nu_k = 1}^\infty
\sum_{p_1,\ldots,p_\nu=0}^\infty g^{2(p_1+\ldots +p_\nu-\nu)}
\left(\
\sum_{\stackrel{m_1,\ldots,m_k}{m = \nu_1m_1+\ldots+\nu_km_k}}
\frac{1}{\nu_1!(m_1!)^{\nu_1}\ldots \nu_k!(m_k!)^{\nu_k}}
 \right.\times\\ \left.\times
\rho^{(p_1|\tilde m_1)}(z_{\sigma(1)},\ldots,z_{\sigma(\tilde m_1)})
\rho^{(p_2|\tilde m_2)}
(z_{\sigma(\tilde m_1+1)},\ldots,z_{\sigma(\tilde m_2)})\ldots
\rho^{(p_\nu|\tilde m_\nu)}
(z_{\sigma(m-\tilde m_\nu+1)},\ldots,z_{\sigma(m)})
\ \right)
\label{With}
\ee

Our task is to express the correlation functions defined
in (\ref{corK}) and (\ref{corrho}) with the help of hat-operators
through the action of check-operators. In fact, this task is already solved
for (\ref{corK}): one can easily see that
\be
\check K^{(\cdot|m)}(z_1,\ldots,z_m)=
\sum_{k_1,...,k_m}{1\over z_1^{k_1+1}...z_m^{k_m+1}}
\check D^{(m)}_{k_1...k_m},\\
K^{(\cdot|m)}(z_1,\ldots,z_m;g) =
Z(T;g)^{-1} \check K^{(\cdot|m)}(z_1,\ldots,z_m) Z(T;g)
\label{corKcheck}
\ee
Moreover, manifest examples that can be found in \cite{amm2} show that
these quantities are expressed through $\check y$ (\ref{checkR}),
its derivatives and $W'(x)$.

Note that one may now invariantly define the quantities that emerged in the
DV solutions, $S_i\equiv {N_i\over g}$. To
this end, one should introduce the ``occupation number" operators (in fact,
these operators describe the monodromy of
$\check K^{(\cdot|1)}(z)$)\footnote{
Relations like
$$
\left[ g^2\oint_{A_i} \check K,\ g^2\oint_{B_j} \check K\right]
\stackrel{?}{\sim}
g^2\left[\check S_i,\frac{\partial}{\partial \check S_j}\right]
= g^2\left[\check N_i,\frac{\partial}{\partial \check N_j}\right]
= g^2\delta_{ij}
$$
should serve as operator counterparts of the celebrated
$$
S_i = \oint_{A_i} \rho^{(0|1)}
$$ $$
\frac{\partial F_{DV}}{\partial S_i} = \oint_{B_i} \rho^{(0|1)}
$$
for the particular DV solution.
However, one should be careful about regularization,
higher-loop corrections etc.
}
\be\label{ocnum}
g\check S_i\equiv {\check N_i} \equiv
\oint_{\alpha_i} \check K^{(\cdot|1)}(z) dz
\ee
Then, $S_i$ are nothing but the eigenvalues of this
operator. This is analogous to the condition
${\p Z\over\p T_0}={N\over g}Z$
which one usually adds to (\ref{vir}) in order to describe matrix integrals.
Now one can define the DV solutions as eigenfunctions of the set of operators
$\check S_i$. Therefore, $S_i$ being eigenvalues, by definition, do not depend
on $T_k$'s. This exactly expresses the isomonodromic properties of the DV
solutions, see, e.g., \cite{Akemann,CM,Ch,Eynard}).

\paragraph{\large Check $\rho$-operators.}

The connected correlators $\rho$ are more "fundamental" than the
full $K$. Therefore, it is natural to wonder if one can find
check-operator analogues of $\rho$'s, once we see that
check-operator counterparts of $K$ do exist and can be of some use.
This means that we would like to put
\be
\check K^{(\cdot|1)}(z;g) =
\sum_{p=0}^\infty g^{2p-2}\check\rho^{(p|1)}(z;g)
\label{Krho1}
\ee
In this way, one gets rid of the terms with explicitly present
operators $\check R$ and prepotentials $F^{(p)}$, the relevant
check-operators $\c \rho^{(\cdot|p)}$ are expressed through $\check y(z;g)$
only (with the single exception of $\check\rho^{(0|1)}(z;g)$,
which also contains $W'(z)$.) Thus, the check-operator $\c K^{(\cdot|p)}$ is a
polynomial in $W'$ of degree $p$. Instead, the $g$ dependence is now distributed between
explicit factors like $g^{2p-2}$ and an additional $g$-dependence
of $\check y(z;g)$. This, however, allows check-operators
$\check \rho^{(p|m)}$ to look exactly the same (modulo ordering)
as the corresponding Gaussian multidensities $\rho_G^{(p|m)}$,
which are all expressed through $y_G$ only.

In \cite{amm2} we suggested {\it a hypothesis} that
{\bf eq.(\ref{Krho1}) is indeed true in all orders in $g^2$ and, moreover,
similar expansions hold for all $\check K^{(\cdot|m)}(z_1,\ldots,z_m;g)$:
they can be all expressed through multilinear combinations
of check operators $\check\rho^{(p|m)}$, which (for $(p|m)\neq (0|1)$)
depend only on $\check y(z;g)$ and its $z$-derivatives in exactly the same
way as $\rho_G^{(p|m)}$ depends on $y_G(z)$}.
However, even to formulate this hypothesis, one needs to introduce
some ordering prescription for products of check-operators
which is not, as usual, unique.
Different ordering prescriptions lead to different explicit
formulas for $\check\rho$, and our hypothesis states that there exist
orderings, when these expressions contain $\check y$, it derivative
and nothing else, except for a few $W'$, see \cite{amm2} for more possible
choices.

In principle, when introducing
$\check\rho$-operators, we have different possibilities of definition, preserving
one or another kind of their relation to $\check K$'s.
They could be defined similarly to (\ref{corKcheck}) from (\ref{corrho}),
so that eq.(\ref{neeq}) below becomes an equality.
However, it appears more interesting {\it instead} of
preserving the equations, to require for $\c \rho^{(\cdot|k)}$
to be the same (up to ordering) as the Gaussian functions $\rho_G^{(\cdot|k)}$.
We can construct recursively an operator modification of expression (\ref{With}).
Since the recurrent equations for $\c K$ are linear
\be
W'(z)K^{(\cdot|m+1)}(z,z_1,\ldots,z_m) -
\check R(z)K^{(\cdot|m)}(z_1,\ldots,z_m) +\\
+\sum_{i=1}^m\frac{\partial}{\partial z_i}
\frac{K^{(\cdot|m)}(z,z_1,\ldots,\check z_i,\ldots, z_m) -
K^{(\cdot|m)}(z_1,\ldots,z_m)}{z-z_i} =
g^2K^{(\cdot|m+2)}(z,z,z_1,\ldots,z_m)
\label{recK}
\ee
they coincide with the equations for
operators $\c K$. The equations for functions $\rho$ are not linear.
Thus, for the operators $\c \rho$ we should choose
the order in which different $\check\rho$ stand in the products.

Note that, with our definition,
\be
\rho_W^{(p|m)}(z_1,\ldots,z_m) \neq
Z(T;g)^{-1} \check\rho_W^{(p|m)}(z_1,\ldots,z_m;g) Z(T;g)
\label{neeq}
\ee
and, in variance with the l.h.s. of (\ref{neeq}),
its r.h.s. is still $g$-dependent.

\paragraph{\large Acknowledgement.}
Our work is partly supported by Federal Program of the Russian Ministry of
Industry, Science and Technology No 40.052.1.1.1112 and by the grants:
RFBR 03-02-17373
and the RFBR grant for support of young scientists,
the grant of the Dynasty foundation and MCFFM  (A.A.),
Volkswagen Stiftung (A.Mir. and A.Mor.),
RFBR 04-01-00646, Grant of Support for the Scientific
Schools 96-15-96798 (A.Mir.), RFBR 04-02-16880 (A.Mor.).

\end{document}